\documentclass[aps,prl,reprint,superscriptaddress,floatfix,amsmath,amssymb]{revtex4-1}

\usepackage{graphicx}
\usepackage{dcolumn}
\usepackage[version=4]{mhchem}
\usepackage{amsmath}
\usepackage{amssymb}
\usepackage{mathrsfs}
\usepackage[T1]{fontenc}

\usepackage[english]{babel}
\usepackage{url}
\usepackage[colorlinks,
linkcolor=blue,
anchorcolor=blue,
urlcolor=blue,
citecolor=blue]{hyperref}

\begin{document}


\title{Thermal Subspace Projection Resolves Particle-Energy Sharing in Entangled Fission}
\author{J.W. Chen}
\affiliation{School of Physics, State Key Laboratory of Nuclear Physics and Technology,
Peking University, Beijing 100871, China}

\author{S.C. Li}
\affiliation{School of Physics, State Key Laboratory of Nuclear Physics and Technology,
Peking University, Beijing 100871, China}

\author{Y. Qiang}
\affiliation{Hangzhou International Innovation Institute of  Beihang University,  Hangzhou 311100, China}
\affiliation{School of Physics, State Key Laboratory of Nuclear Physics and Technology,
Peking University, Beijing 100871, China}

\author{J.C. Pei}\email{peij@pku.edu.cn}
\affiliation{School of Physics, State Key Laboratory of Nuclear Physics and Technology,
Peking University, Beijing 100871, China}
\affiliation{Southern Center for Nuclear-Science Theory (SCNT), Institute of Modern Physics, Chinese Academy of Sciences, Huizhou 516000, China}

\date{\today}

\begin{abstract}
Nuclear fission realizes a spatially bipartite entangled many-body system in which well-separated fragments counterintuitively share particles and energy.
How this quantum sharing mechanism behaves at finite excitation energy is an outstanding question for induced fission.
We formulate, for the first time, a thermal double particle-number projection onto spatial subspaces, using the Balian-Brezin representation and Grassmann integrals, thereby deriving the projected partition function and fragment observables.
Applied to finite-temperature dynamical fission calculations of \(^{240}\)Pu, the formalism provides a microscopic explanation for
the puzzling energy dependence of odd-even staggering in charge yields and sawtooth structures in neutron multiplicities, shedding new light on the quantum nature of nuclear scission.
\end{abstract}

\maketitle

\emph{Introduction.}---Nuclear fission presents a rare natural example of spatially bipartite entanglement in many-body systems.
Counterintuitively, the persistent entanglement alters the sharing of particles and energies between well-separated fragments~\cite{qiangQuantumEntanglementNuclear2025},
exhibiting a matter-energy teleportation virtually.
This contrasts with the longstanding classical view of nuclear scission as a random neck rupture~\cite{brosa}.
Particularly intriguing is how the quantum sharing mechanism evolves with excitation energy under the competing influences of entanglement and thermal effects.
This question is central to induced fission and the energy dependence of post-scission observables, which
poses a major challenge for nuclear data needs~\cite{nd,endf}.

Neutron-induced fission is vital to nuclear energy production and interdisciplinary applications; however,
both experiments and fission theories
fall well short of the precision needed~\cite{schunckMicroscopicTheoryNuclear2016,andreyev_nuclear_2017,bender_future_2020,schunckTheoryNuclearFission2022}.
Experimentally the fission data between thermal neutron energy 0.0253 eV and 14 MeV are rather sparse and imperfect\cite{gooden_energy_2016,gooden_energy_2024,kelly_measurement_2020},
hindering the energy interpolation of key observables such as fission product yields and neutron multiplicities--observables that probe particle and energy sharing, respectively.
In particular, it is highly desirable to describe the attenuation of odd-even staggering in charge yield distributions~\cite{pomme_excitation_1993,chatillonExperimentalStudyNuclear2019,ramosInsightExcitationEnergy2019a,verriereMicroscopicCalculationFission2021,ramos_experimental_2023,mollerDetailedModelingOdd2024,huangRoleTensorForce2024,liu_physics_2026}
and the sawtooth structures in neutron multiplicities~\cite{naqvi_fission_1986,thulliezNeutrongammaMultiplicities2019,albertssonExcitationEnergyPartition2020}  as excitation energy increases.
Theoretically, microscopic fission dynamics are predominantly applied to  spontaneous fission or fission at zero temperature~\cite{bulgacInducedFissionPu2016a,scamps_superfluid_2015,tanimura_microscopic_2017,goutte_microscopic_2005,regnier_fission_2016,pear, goddard_fission_2015,umar}.
The induced fission at finite excitation energies~\cite{qiangEnergyPairingDependence2021,qiang2021,zhao_microscopic_2019} is far more complex and remains the least understood.

Entanglement in fission corresponds to a superposition of configurations with different nucleon-number partitions between two complementary fragments.
The double particle number projection (DPNP)  onto the subspace  has been proposed to extract nucleon-number-resolved yields of different binary combinations~\cite{verriereNumberParticlesFission2019,verriereMicroscopicCalculationFission2021}.
The entanglement magnitude is strongly correlated with the variations of particle numbers in subsystems, and thus the
distribution widths of fission yields~\cite{li_spatial_2026}.
At finite excitations, the distribution widths are determined by the competition between the fading entanglement and the rising thermal broadening.
Along with the teleportation of particles, energy is redistributed,
impacting the energy sharing which can also be described within the double projection framework~\cite{qiangQuantumEntanglementNuclear2025,sekizawaParticlenumberProjectionMethod2014,shangEnergyPartitionSplitting2025}.

DPNP has so far been applied to splitting wave functions from time-dependent density functional theory (TDDFT) only at zero temperature~\cite{simenelParticleTransferReactions2010,scampsEffectPairingOne2013,sekizawaParticlenumberProjectionMethod2014,verriereNumberParticlesFission2019,qiangQuantumEntanglementNuclear2025,shangEnergyPartitionSplitting2025,robledoParticleNumberProjection2025}. The resulting distribution widths of
fragment products due to entanglement are about 4  mass units~\cite{li_spatial_2026}, a considerable fraction of the total distribution.
However, DPNP at finite excitations  has not been realized yet,
although finite-temperature particle number projection on the full space after variation has been developed~\cite{esebbagNumberProjectedStatistics1993,rossignoliProjectionFiniteTemperature1994,tanabeQuantumNumberProjection2005,fantoParticlenumberProjectionFinitetemperature2017,
chenParticleNumberProjected2026}.
The thermal full-space projection restores the particle numbers violated by both BCS pairing correlations and the grand canonical ensemble.
For spatial subspace projection, the nonunitary character of spatially restricted gauge transformations is inherent, irrespective of pairing correlations, as pointed out by L. Robledo~\cite{robledoParticleNumberProjection2025}.
Particle sharing can be extracted from the projected partition function, whereas calculating fragment observables is more difficult owing to the noncommutativity between the observable and the subspace projection operator.

In this Letter, we formulate finite-temperature DPNP onto spatial subsystems to unravel entangled fission at finite excitation energies.
The projection is applied to wave functions from time-dependent density functional theory (TDDFT) calculations of $^{240}$Pu at finite temperature~\cite{qiang2021}.
The thermal subspace projection is formulated by leveraging the Grassmann integral method for many-fermion quantum systems.
Using the Balian--Brezin decomposition \cite{balianNonunitaryBogoliubovTransformations1969}, and building on Pfaffian treatments of the Hartree-Fock-Bogoliubov (HFB) overlap sign ambiguity \cite{robledoSignOverlapHartreeFockBogoliubov2009,bertschSymmetryRestorationHartreeFockBogoliubov2012}, we express the integral kernel in the projected partition function as a Pfaffian whose complex phase is obtained directly.
We derive the associated transition normal and anomalous densities, which allow projected fragment observables to be evaluated with the finite-temperature generalized Wick theorem~\cite{piresGeneralizedWicksTheorem2012}.
The theoretical framework provides the first self-consistent microscopic descriptions of  energy dependencies of fission yield distributions and fragment excitation energies, addressing
the puzzling odd-even structures in charge yields~\cite{pomme_excitation_1993,chatillonExperimentalStudyNuclear2019,ramosInsightExcitationEnergy2019a,verriereMicroscopicCalculationFission2021,ramos_experimental_2023,mollerDetailedModelingOdd2024,huangRoleTensorForce2024,liu_physics_2026} and the sawtooth structures in neutron multiplicities~\cite{naqvi_fission_1986,thulliezNeutrongammaMultiplicities2019,albertssonExcitationEnergyPartition2020}.
To demonstrate the quantum-thermal interplay, we also perform systematic projection calculations with fixed-background wave functions.

\emph{Thermal Subspace Projection.} ---
As the fissioning nucleus splits into two complementary fragments, DPNP is performed onto two
 spatial subspaces $L$ and $H$ simultaneously. The nucleon numbers of two fragments are resolved with their sum conserved.
In the grand canonical ensemble, the corresponding projected partition function is given by~\cite{sheikhSymmetryRestorationMeanfield2021}
\begin{align}
    Z_{P}=\frac{1}{4\mathrm{\pi}^2}\int_{0}^{2\mathrm{\pi}}\mathrm{d}\theta_{{L}}\mathrm{d}\theta_{\text{H}}\,\mathrm{e}^{-\mathrm{i}\theta_{\text{L}} N_{\text{L}}}\mathrm{e}^{-\mathrm{i}\theta_{\text{H}} N_{\text{H}}}\tilde{Z}(\theta_{\text{L}}, \theta_{\text{H}}),
    \label{projz}
\end{align}
where $\tilde{Z}(\theta_{\text{L}}, \theta_{\text{H}})=\operatorname{Tr}(\hat{D}\mathrm{e}^{\mathrm{i}\theta_{\text{L}}\hat{N}_{\text{L}}}\mathrm{e}^{\mathrm{i}\theta_{\text{H}}\hat{N}_{\text{H}}})$ is referred to as the generating function; $\hat{N}_{\rm L/H}$ denotes the particle number operator in subspaces.
In an orthonormal single-particle basis, $\hat{N}_{\rm L/H}=\sum_{ij}O^{}_{L/H}a_{i}^{\dagger}a_{j}$, with $O_{L/H}=\langle i|\Theta_{\rm L/H}(\boldsymbol{r})|j\rangle$, and $\Theta_{\rm L/H}$ is the spatial indicator function.
Note that the formalism of $Z_P$ is connected to the full counting statistics in mesoscopic physics and quantum transport~\cite{lacroixCountingStatisticsFinite2020,levitov_electron_1996,kindermann_full_2003,esposito_nonequilibrium_2009}.
Writing the HFB Hamiltonian as a scalar part $H_0$ plus a quadratic form~\cite{alanl.goodmanFinitetemperatureHFBTheory1981}, the unnormalized statistical operator can be decomposed as $\hat{D}=\mathrm{e}^{-\beta H_0}\hat{\mathcal D}$, where the ensemble is at temperature $T=1/\beta$ and the Balian-Brezin~\cite{balianNonunitaryBogoliubovTransformations1969} representation gives
\begin{align}
    \hat{\mathcal D}=\mathrm{e}^{\frac{1}{2}\boldsymbol{a}^{\dagger}X\boldsymbol{a}^{\dagger}}\mathrm{e}^{\boldsymbol{a}^{\dagger}Y\boldsymbol{a}}\mathrm{e}^{\frac{1}{2}\boldsymbol{a}Z\boldsymbol{a}},
\end{align}
The matrices $X$, $Y$, and $Z$ can be obtained explicitly in the BCS canonical basis~\cite{ebataCanonicalbasisTimedependentHartreeFockBogoliubov2010}, as shown in the Supplemental Material~\cite{supplementalMaterial}.

Using the fermionic coherent-state formulation, the generating function in Eq.(\ref{projz}) becomes
\begin{align}
    \tilde{Z} = \mathrm{e}^{-\beta H_0}\int\mathrm{d}\mu(\boldsymbol{z})\,\mathrm{e}^{-\frac{1}{2}\operatorname{Tr}(Y)}
    \langle-\boldsymbol{z}|\hat{\mathcal D}\hat{K}|\boldsymbol{z}\rangle,
\end{align}
where $\hat{K}(\theta_{\text{L}}, \theta_{\text{H}})=\mathrm{e}^{\mathrm{i}\theta_{\text{L}}\hat{N}_{\text{L}}}\mathrm{e}^{\mathrm{i}\theta_{\text{H}}\hat{N}_{\text{H}}}$ is the kernel operator, and its corresponding matrix $K$ is $\mathrm{e}^{\mathrm{i}\theta_{\text{L}}O_{\text{L}}}\mathrm{e}^{\mathrm{i}\theta_{\text{H}}O_{\text{H}}}$.
Introducing $\zeta_{\mu}=\begin{bmatrix}\boldsymbol{z}^{*} & \boldsymbol{z}\end{bmatrix}_{\mu}$ and the antisymmetric matrix
\begin{align}
    \mathbb{M}=\begin{bmatrix}X & -[\mathrm{e}^{Y}K+1]\\
    [\mathrm{e}^{Y}K+1]^{\mathrm{T}} & K^{\mathrm{T}}ZK
\end{bmatrix},
\label{mm}
\end{align}
the generating function can then be written as the Grassmann integral
\begin{align}
    \tilde{Z} = s_{N}\mathrm{e}^{-\beta H_0}\int\mathrm{d}\boldsymbol{\zeta}\,\mathrm{e}^{-\frac{1}{2}\operatorname{Tr}(Y)}\mathrm{e}^{\frac{1}{2}\boldsymbol{\zeta}^{\mathrm{T}}\mathbb{M}\boldsymbol{\zeta}},
\end{align}
where $s_{N}=(-1)^{N(N-1)/2}$ and  $N$ here denotes the single-particle
basis dimension. Evaluating the Gaussian Grassmann integral gives
\begin{align}
    \tilde{Z} = s_{N}\mathrm{e}^{-\beta H_0}\mathrm{e}^{-\frac{1}{2}\operatorname{Tr}(Y)}\operatorname{pf}(\mathbb{M}).
\end{align}

For the thermal HFB statistical operator, both $\mathrm{e}^{-\beta H_0}$ and $\mathrm{e}^{-\frac{1}{2}\operatorname{Tr}(Y)}$ are positive and real, so the complex phase is determined unambiguously by the Pfaffian.
We note that a different formulation of partition function was derived by P. Fanto~\cite{p.fantoProjectionVariationFinitetemperature2017}, which involves additional treatments of the complex phases.

\begin{figure*}[!t]
    \centering
    \includegraphics[width=\textwidth]{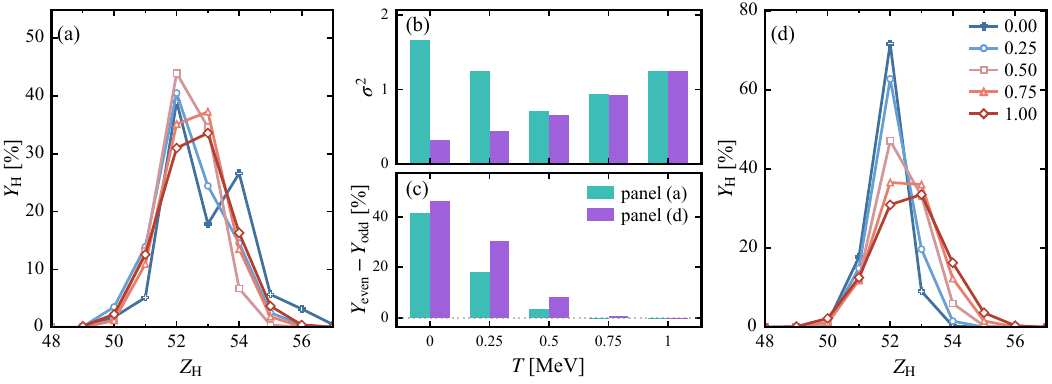}
    \caption{ Fragment charge yields from thermal subspace projections of  $^{240}$Pu at different temperatures.
    (a) the self-consistent projection results of charge yields $Y_Z$ for heavy fragments.
    (b) variations $\sigma_Z^2$ of proton numbers in charge yields of heavy fragments from self-consistent and fixed-background calculations, corresponding to panel(a) and panel(d), respectively.
    (c) the difference between fragment yields of even and odd charge numbers from self-consistent and fixed background calculations.
    (d) the distributions of charge yields from thermal subspace projections with fixed backgrounds, in which wave functions, single-particle Hamiltonian and pairing fields are held at $T$=1.0 MeV but their thermal occupations change from $T=0$ to $1.0$ MeV.}
    \label{fig:fragment-number-distributions}
\end{figure*}

For an observable $\hat{O}$ that commutes with the projector, its projected expectation value $\langle\hat{O}\rangle$ is
\begin{align}
    \langle\hat{O}\rangle=\frac{1}{Z_{P}}\frac{1}{4\mathrm{\pi}^2}\int\mathrm{d}\boldsymbol{\theta}\,\mathrm{e}^{-\mathrm{i}\boldsymbol{\theta}\cdot\boldsymbol{N}}\operatorname{Tr}(\hat{D}\hat{K}\hat{O}),
\label{po}
\end{align}
where $\boldsymbol{\theta}=(\theta_{\text{L}}, \theta_{\text{H}})$ and $\boldsymbol{N}=(N_{\text{L}}, N_{\text{H}})$ are shorthand notations. In general, $\hat{N}_{L/H}$ does not commute with the Hamiltonian because particles can cross the scission boundary.
The thermal full-space projection can not be directly applied since the observable does not commute with the projector.
For well-separated fission fragments, the associated boundary terms are negligible, so the projected fragment energy can be evaluated under the separated-fragment condition.

We define the normalized gauge-angle contraction as $\langle\hat{O}\rangle_{\boldsymbol{\theta}}=\operatorname{Tr}(\hat{D}\hat{K}\hat{O})/\tilde Z(\boldsymbol{\theta})$.
Because $\hat{D}\hat{K}$ is Gaussian, the finite-temperature generalized Wick theorem remains valid \cite{piresGeneralizedWicksTheorem2012}.
Any observable can therefore be expressed in terms of the transition contractions $\tilde{\rho}_{qp}=\langle a_{p}^{\dagger}a_{q}\rangle_{\boldsymbol{\theta}}$, $\tilde{\kappa}_{qp}=\langle a_{p}a_{q}\rangle_{\boldsymbol{\theta}}$, and $\tilde{\lambda}_{qp}=\langle a_{q}^{\dagger}a_{p}^{\dagger}\rangle_{\boldsymbol{\theta}}$.
Using $\mathrm{e}^{\mathrm{i}\theta\hat{N}_{L/H}}a^{\dagger}_{i}\mathrm{e}^{-\mathrm{i}\theta\hat{N}_{L/H}}=\sum_{j}[\mathrm{e}^{\mathrm{i}\theta O}]_{ji}a^{\dagger}_{j}$, the normal contraction is
\begin{align}
    \tilde{\rho}_{qp} =\delta_{pq}-\sum_{r}K^{\dagger}_{rp}\frac{\operatorname{Tr}(a_{r}^{\dagger}\hat{D}a_{q}\hat{K})}{\operatorname{Tr}(\hat{D}\hat{K})}.
    \label{eq:transition-rho}
\end{align}

The numerator $\operatorname{Tr}(a_{r}^{\dagger}\hat{D}a_{q}\hat{K})$ can be evaluated in the coherent-state representation
and written as
\begin{align}
     \mathrm{e}^{-\beta H_0}\int\mathrm{d}\mu(\boldsymbol{z})\mathrm{e}^{-\frac{1}{2}\operatorname{Tr}(Y)}\langle-\boldsymbol{z}|a_{r}^{\dagger}\hat{\mathcal D}a_{q}\hat{K}|\boldsymbol{z}\rangle.
\end{align}

This expression reduces to the Grassmann integral
\begin{align}
     s_{N}\mathrm{e}^{-\beta H_0}\sum_{s}K_{qs}\mathrm{e}^{-\frac{1}{2}\operatorname{Tr}(Y)}\int\mathrm{d}\boldsymbol{\zeta}\,z_{r}^{*}z_{s}\mathrm{e}^{\frac{1}{2}\boldsymbol{\zeta}^{\mathrm{T}}\mathbb{M}\boldsymbol{\zeta}}.
\end{align}

The anomalous contractions follow analogously. Finally the three transition contractions are derived as
\begin{equation}
\begin{aligned}
    \tilde{\rho} & =[1-K^{*}\mathbb{M}^{-1}_{12}K^{\mathrm{T}}]^{\mathrm{T}},\\
    \tilde{\kappa} & =-K\mathbb{M}_{22}^{-1}K^{\mathrm{T}},\\
    \tilde{\lambda} & =K^{*}\mathbb{M}_{11}^{-1}K^{\dagger},\\
\end{aligned}
\label{tdd}
\end{equation}
where $\mathbb{M}^{-1}_{11}$, $\mathbb{M}^{-1}_{12}$ and  $\mathbb{M}^{-1}_{22}$ denote the $2\times2$ blocks of $\mathbb{M}^{-1}$,
corresponding to the matrix $\mathbb{M}$ in Eq.(\ref{mm}).
Each contraction is evaluated at the same gauge angles as the
trace and then enters the projected observables in Eq.(\ref{po}).
The binding energies of fragments can be obtained using transition densities from these three contractions.
More details of the theoretical framework, including dynamical fission calculations, the canonical basis transformation of the operators, validations of thermal subspace projection at limits of thermal full-space projection or zero-temperature subspace projection,
and the separated-fragment condition for  Eq.(\ref{po}) are given in the Supplemental Material~\cite{supplementalMaterial}.

\emph{Particle Sharing.}---
The thermal  DPNP onto
wave functions of compound fissioning nuclei is performed, and
then the proton number $Z$ and the
neutron  number $N$ of both light and heavy fragments are resolved.
Projection onto both subspaces is equivalent to projection onto either subspace plus the full space.
Since the obtained probabilities of two complementary fragments are the same, we only show those of the heavy fragments.
The yield or probability of a specific fragment is given by
 \begin{equation}
 Y_H(N_H,Z_H)=Z_P(N_H,Z_H,N,Z)/Z_P(N,Z),
 \end{equation}
where $Z_P(N,Z)$ is the projected partition function in the full space.
The normalization of fragment yields is guaranteed.
The charge yield distribution is the marginal $Y_H(Z_H)=\sum_{N_H}Y_H(N_H,Z_H)$.
The dynamical fission trajectory is obtained with the Sky3D solver~\cite{sky3d,sky3d2}, adopting initial inputs from deformation-constrained static calculations at finite temperature $T$ using the SkyAx solver~\cite{reinhardAxialHartreeFock2021}. The SkM$^{*}$~\cite{skm} nuclear force and the density dependent pairing interaction~\cite{mixpair} are used as in our previous work~\cite{qiang2021}.
Here $T=1/\beta$ is a control parameter associated with initial excitation energies $E^{*}$ of compound nuclei and the ensemble.
In self-consistent calculations, the same temperature is employed in the fission process and the thermal subspace projection.

\begin{figure*}[htbp]
    \centering
    \includegraphics[width=\linewidth]{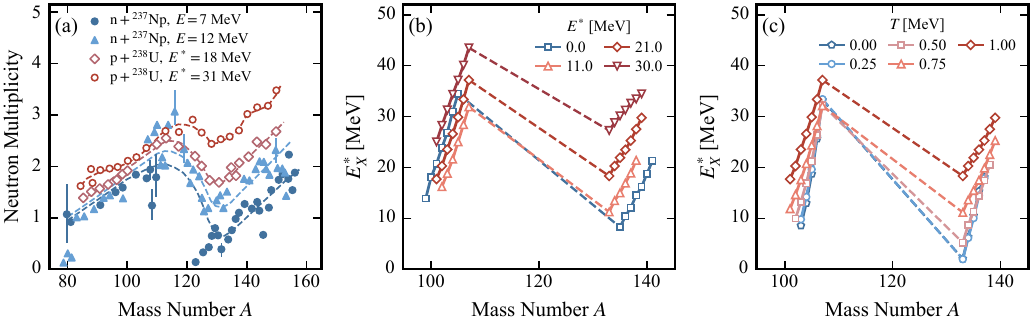}
    \caption{Fragment excitation energies $E_X^{*}$ of  $^{240}$Pu as a function of fragment mass number.
    (a) the sawtooth structures of experimental average neutron multiplicities from fission of compound nuclei $^{238,239}$Np~\cite{defrenneChargeDistributionsPhotofission1982,u238p}, relating to fragment excitation energies.
    (b) self-consistent thermal subspace projection results from $E^{*}$=0 to 30 MeV .
    (c) projected results with fixed backgrounds, in which wave functions and Hamiltonian are held at $T$=1.0 MeV but their
    thermal occupations changes at varying temperatures. }
    \label{fig:fragment-excitation-temperature}
\end{figure*}

Fig.~\ref{fig:fragment-number-distributions} shows the projected charge yield distributions of $^{240}$Pu at different temperatures, relating to the sharing of particles between entangled fragments at finite excitations.
Generally the distributions are rather narrow since only the post-scission sharing are considered presently within the TDDFT framework.
Fig.~\ref{fig:fragment-number-distributions}(a) shows the charge yields with temperatures $T$ ranging from 0.0 to 1.0 MeV, obtained self-consistently.
To extract the thermal effect,  Fig.~\ref{fig:fragment-number-distributions}(d) shows projection results with fixed background  in which wave functions and Hamiltonian are
held at $T$=1.0 MeV, but their thermal occupations  change at varying temperatures.
Note that fission dynamical calculations cannot reach scission at higher temperatures without invoking dynamical fluctuations~\cite{qiang2021}.

With increasing temperatures, the obtained particle number variances $\sigma_Z^2$, related to distribution widths, are shown in Fig.~\ref{fig:fragment-number-distributions}(b).
Results show that the charge yield distribution widths  decrease with increasing temperatures at the beginning, and
then increase after $T$=0.5 MeV.
It is known that the distribution width  at zero temperature is correlated with the entanglement magnitude~\cite{li_spatial_2026}.
First, the fading of quantum coherent effects such as entanglement and pairing correlations leads to reduced distribution widths.
Then, increasing thermal broadening effects come into play and increase the distribution widths.
The evolution of yield widths exhibits an interesting competition between quantum and thermal effects.
In the fixed-background projection, the variances of yield distributions increase steadily as
$T$ increases.
This thermal broadening effect is attributed to the mixing of high-excitation configurations in the ensemble.
In experiments, the widths of fission yields are indeed broader at high excitations~\cite{atreya_fragment-mass_2024,sen_fission_2017,hambsch_study_2000}.
The fixed-background projection shows that the particle-number variances, which reflect the thermal broadening effect, are approximately linearly correlated with $T$.
Besides, the projected distributions become more and more like a Gaussian function, a situation similar to the thermal full-space projection~\cite{chenParticleNumberProjected2026}.

The odd-even staggering in charge yields is a  characteristic  fine structure and inferring its energy dependence is a long-standing issue~\cite{pomme_excitation_1993,chatillonExperimentalStudyNuclear2019,ramosInsightExcitationEnergy2019a,verriereMicroscopicCalculationFission2021,ramos_experimental_2023,mollerDetailedModelingOdd2024,huangRoleTensorForce2024,liu_physics_2026}.
The odd-even structure is clearly shown at zero temperature in Fig.~\ref{fig:fragment-number-distributions}(a).
To quantify the odd-even staggering magnitude, Fig.~\ref{fig:fragment-number-distributions}(c) shows the subtracted fission yields between even and odd charge numbers.
The odd-even structure quickly decreases with increasing temperatures and  disappears at $T$=0.75 MeV.
In the projection with fixed backgrounds, the odd-even effect is slightly larger and disappears later.
In both cases, the odd-even structures decrease almost linearly with increasing $T$ rather
than excitation energies. This provides a useful relation for guiding phenomenological models.
The odd-even structures are almost washed out at $T$=0.75 MeV, corresponding to an excitation energy $E^{*}$=11.0 MeV.
For comparison, the odd-even staggering persists above $E^{*}$=16 MeV based on the time-dependent generator-coordinate method with DPNP at zero temperature~\cite{verriereMicroscopicCalculationFission2021}.
A recent experiment on $^{240}$Pu shows weak odd-even effects at ${E^{*}}$=8.2 MeV, which totally disappear at $E^{*}=11.9\,\text{MeV}$~\cite{ramos_isotopic_2026}.
This demonstrates the essential role of thermal subspace projection for the attenuation of odd-even structures in charge yields. 
The calculated odd-even structures in terms of neutron numbers are not prominent at all and not shown.

\emph{Energy Sharing.}---
The excitation energies of fragments are directly related to neutron multiplicities~\cite{tudoraRevisitingResidualTemperature2018,brosaSawtoothCurveNeutron1985,thulliezNeutrongammaMultiplicities2019,talou_correlated_2018}, as a probe of energy sharing at finite excitation energies.
The binding energies of fragments can be obtained by integrating over the energy density functional at each gauge angle $\theta$ in the statistical ensemble,
\begin{align}
E_{\rm X} & =\frac{1}{Z_{P}}\frac{1}{4\mathrm{\pi}^{2}}\int\mathrm{d}\boldsymbol{\theta}\mathrm{e}^{-\mathrm{i}
\boldsymbol{\theta}\cdot\boldsymbol{N}}\mathop{\text{Tr}}(\hat{D}\hat{K}(\boldsymbol{\theta}))E(\boldsymbol{\theta})
\end{align}
where the transition densities $\tilde{\rho}$, $\tilde{\kappa}$, $\tilde{\lambda}$ in Eq.(\ref{tdd}) are used, and $E(\boldsymbol{\theta})$ denotes the energy density functional as given in~\cite{qiangQuantumEntanglementNuclear2025} including pairing energies,
which are constructed on these transition densities.
The excitation energies of fragments with a specific ($Z_{\rm X}$,$N_{\rm X}$) can be evaluated by subtracting the center-of-mass kinetic energies and ground state energy $E_{gs}$ as
\begin{equation}
    E_X^*=E_{X}-\frac{\hbar^2\left|\langle\hat{\boldsymbol{P}}_X\rangle\right|^2}{2mA_X}-E_{\mathrm{gs}}(A_X,Z_X),
\label{mpx}
\end{equation}
where $A_X=N_X+Z_X$ is the fragment mass number.
It has to be mentioned that the momentum $\langle\hat{\boldsymbol{P}}_X\rangle$ is also given by the thermal subspace projection onto specific fragments.
$E_{\mathrm{gs}}$ is the ground-state energy of the corresponding fragment calculated with the same Skyrme force.

Fig.~\ref{fig:fragment-excitation-temperature} shows how the fragment excitation energies evolve with mass number and temperatures.
For each mass number, the excitation energy $E_X^{*}$ displayed  is the probability-weighted average over all nucleon-number combinations,
and $E_X^{*}$ is proportional to neutron multiplicities.
Fig.~\ref{fig:fragment-excitation-temperature}(a) shows  experimental average neutron multiplicities of compound nuclei $^{238, 239}$Np~\cite{defrenneChargeDistributionsPhotofission1982,u238p},
exhibiting the fascinating sawtooth structures up to $E^{*}$=31 MeV.
Fig.~\ref{fig:fragment-excitation-temperature}(b)
 presents the self-consistent results from $T=0$ and $1.2$ MeV, corresponding to $E^{*}$=0 to 30 MeV.
Note that by employing the  projected momentum $P_X$ in Eq.(\ref{mpx}),  the sawtooth dependencies on mass numbers are substantially reduced.
The is crucial to obtain reasonable neutron multiplicities, while the average momentum without projection was adopted before~\cite{sekizawaParticlenumberProjectionMethod2014,qiangQuantumEntanglementNuclear2025}.
To drive dynamical evolutions at $T$=1.2 MeV, a small initial boost is employed~\cite{sky3d}, for which
the total excitation energy is 30 MeV.
For light fragments, $E_X^{*}$ and their slopes
  do not change much with increasing excitation energies. In contrast, $E_X^{*}$
  for heavy fragments increases steadily, while the sawtooth slope remains almost unchanged up to 11 MeV.
  For $E^{*}$=21 to 30 MeV, the sawtooth slopes for heavy fragments decrease slightly.
For the first time, the behavior of energy dependent sawtooth of neutron multiplicities is explained microscopically.
Existing fission models have
 long assumed different temperatures for light and heavy fragments to explain the sawtooth structures~\cite{entropy}.
 However, in statistical descriptions of energy sharing, the sawtooth structures disappear already at $E^{*}$=20 MeV~\cite{albertssonExcitationEnergyPartition2020},
 which is considerably lower than expected.

Fig.~\ref{fig:fragment-excitation-temperature}(c) displays the projected excitation energies of fragments
with fixed backgrounds at $T$=1.0 MeV, corresponding to fission yields in Fig.~\ref{fig:fragment-number-distributions}(d).
 It shows that the excitation energies of both light and heavy fragments increase with temperature, but the increase is faster for heavy fragments.
 Moreover, the sawtooth slopes decrease steadily, with the change being more pronounced for heavy fragments.
In contrast,  Fig.~\ref{fig:fragment-excitation-temperature}(b) shows that the slopes for light fragments remain  unchanged,
whereas those for heavy fragments decrease much more slowly.
 This comparison underscores the role of self-consistent structural rearrangement.
The self-consistent trend implies the persistence of sawtooth structures even at considerable excitation energies,
characterized by steadily increasing excitation energies and slowly decreasing sawtooth slopes for heavy fragments.
Presently we focus on the post-scission sharing of particles and energies, nevertheless, the pre-scission sharing mechanism including
the fragment deformation effects and shell effects
are also essential~\cite{albertssonExcitationEnergyPartition2020}.

\emph{Conclusion.}---In this Letter,  we present the first formulation of  thermal double particle-number projection for spatially entangled subsystems
 to probe the counterintuitive sharing of particles and energies between separated fission fragments at finite excitation energies.
 Our method is based on the Balian-Brezin decomposition and  Grassmann integral, in which  a phase-unambiguous Pfaffian gauge kernel and associated transition normal and anomalous densities are derived.
The formalism can revert to the limits of zero-temperature subspace projection and finite-temperature full-space projection.

When applied to induced fission of $^{240}$Pu,  the puzzling features of odd-even staggering and sawtooth structures
are explained microscopically, highlighting  the quantum nature of nuclear scission. The resulting energy dependence exhibits an interesting interplay between quantum and thermal effects
and provides useful guidance for phenomenological fission models.
Furthermore, self-consistent calculations demonstrate that such sawtooth structures persist at considerable excitation energies as expected.

This framework provides a practical foundation for inferring observables in neutron-induced fission that are notoriously difficult to measure experimentally. These methods are also applicable to multi-nucleon transfer reactions at high excitation energies. More broadly, thermal subspace projection offers a general tool for probing spatially nonlocal transport and entanglement in thermal many-body systems.

 \acknowledgments
 We thank F.R. Xu for useful discussions.
 This work was supported by the National Key R$\&$D Program of China (Grant No.2023YFA1606403) and the National Natural Science Foundation of China under Grants Nos.12475118 and 12335007.


\end{document}